\documentclass[useAMS,usenatbib,fleqn]{mnras}
\usepackage{graphicx,color,amsmath,amssymb,amsfonts}
\usepackage{subfig}
\usepackage[toc,page]{appendix}
\usepackage{hyperref} 
\usepackage{etoolbox}
\usepackage{natbib}

\newcommand{\be}{\begin{equation}}
\newcommand{\ee}{\end{equation}}
\newcommand{\bea}{\begin{eqnarray}}
\newcommand{\eea}{\end{eqnarray}}

\usepackage{graphicx,pdflscape}

\title[A new sample of southern radio galaxies]{A new sample of southern radio galaxies: Host galaxy masses and star-formation rates}

\author[Marubini et al.]{Takalani Marubini$^{1}$\thanks{E-mail: takalanimarubini@gmail.com},
Matt J.~Jarvis$^{1,2}$,
Stephen Fine$^{1}$,
Tom Mauch$^{3}$, \and
Kim McAlpine$^{1,3}$,
Matthew Prescott$^{1}$
\\
$^{1}$Department of Physics \& Astronomy, University of the Western Cape, Private Bag X17, Bellville 7535, South Africa\\
$^{2}$Astrophysics, University of Oxford, Denys Wilkinson Building, Keble Road, Oxford, OX1 3RH, United Kingdom\\
$^{3}$South African Radio Astronomy Observatory (SARAO), 2 Fir Street, Black River Park, Observatory, 7925, South Africa
}

\date{Accepted XXX. Received YYY; in original form ZZZ}

\pubyear{2019}

\begin{document}
\label{firstpage}
\pagerange{\pageref{firstpage}--\pageref{lastpage}}
\maketitle

\begin{abstract}

In this study we define a new sample of distant powerful radio galaxies in order to study their host-galaxy properties and provide targets for future observations of H{\sc i} absorption with new radio telescopes and to understand the fuelling and feedback from such sources.
We have cross-matched the Sydney University Molonglo Sky Survey 
(SUMSS) radio catalogue at 843\,MHz with the VISTA Hemisphere Survey (VHS) near-infrared catalogue using the Likelihood Ratio technique. Photometric redshifts from the Dark Energy Survey are then used to assign redshifts to the radio source counterparts.  We found a total of 249 radio sources with
photometric redshifts over a 148\,deg$^2$ region.  By fitting the optical and near-infrared photometry with spectral synthesis models we determine the stellar mass and star-formation rates of the radio sources, finding typical stellar masses of $10^{11} - 10^{12}$\,M$_{\odot}$ for the powerful high-redshift radio galaxies. We also find a population of low-mass blue galaxies. However, by comparing the derived star-formation rates to the radio luminosity, we suggest that these sources are  false positives in our likelihood ratio analysis.
Follow up, higher-resolution ($\lesssim 5$\,arcsec) radio imaging would help alleviate these mid-identifications, as the limiting factor in our cross-identifications is the low resolution ($\sim 45$\,arcsec) of the SUMSS radio imaging.

\end{abstract}




\section{Introduction}

Powerful radio galaxies and radio-loud quasars ($L_{\rm 1.4GHz} \gtrsim 10^{25}$\,W\,Hz$^{-1}$) form an important subset of the general Active Galactic Nuclei (AGN) population. Their large-scale radio emission permits them to be selected at radio wavelengths where dust-obscuration is unimportant, thus plausibly allowing a census of a population to be made \citep[e.g.][]{Willott01,McAlpine13,Smolcic2017c}. Furthermore, over the past decade, radio galaxies have become a key component in simulations and models of the formation and evolution of massive galaxies. The radiative emission from the high-accretion rate sources can heat up the gas in and around the host galaxy, and the powerful radio jets that emanate from such galaxies, along with their lower-accretion rate counterparts can provide a form of feedback, due to the mechanical work that the jets can do against the inter-stellar and inter-galactic media \citep[e.g.][]{RawlingsJarvis2004,Croton2006,Bower2006,Schaye2015,Sijacki2015,Beckman2017}..

For many decades, the best studied sample of radio galaxies was derived from the third Cambridge (3C) survey, and the revised version \citep[3CRR;][]{LRL83}. A large observational campaign to obtain spectroscopic redshifts for these sources provided the benchmark for detailed studies of powerful radio galaxies across all wavelengths \citep[e.g.][]{LillyLongair1984,Best1997,Willott03}. However, although the basis for a huge amount of studies, 3CRR is a single flux-density limited sample and as such is subject to Malmquist bias. For this reason, observational campaigns towards the end of the 20th Century concentrated on obtaining complete redshift information for progressively fainter radio surveys \citep[e.g.][]{Rawlings2001,Willott2002,Jarvis2001a}.

These surveys were the bedrock for gaining an understanding the physics of powerful radio-loud AGN \citep[e.g.][]{Kaiser1997,Hardcastle1998,Blundell1999}, their space density evolution \citep[e.g.][]{Willott01,Jarvis2000,Jarvis2001c,Cruz2007} and their host galaxies across cosmic time. Firstly using the observed relationship between the near-infrared $K-$band magnitude and redshift \citep[the radio galaxy $K-z$ relation, e.g.][]{LillyLongair1984,Eales1997,Jarvis2001b,Willott03}, and subsequently through deep {\em Spitzer} observations \citep[e.g.][]{Seymour2007} and multi-band photometry \citep[e.g.][]{Fernandes2015}. 
In recent years, there has been a shift in focus towards the less luminous population of radio sources, for which the evolution is less well understood \citep[e.g.][]{ClewleyJarvis2004,Sadler2007,McAlpine13,Rigby2015,Smolcic2017c}. Such studies require deeper radio data, whilst retaining the ability to measure redshifts for the radio sources. This has led these studies to rely predominantly on photometric redshifts, but see \cite{Prescott2016} and \cite{Pracy2016}, who make use of spectroscopy.

Creating new samples of powerful radio sources with redshift information is still important. This is because they are intrinsically rare and even the spectroscopically complete 3CRR, 6C and 7CRS samples only contain a few hundred objects. Our ability to create new samples of powerful AGN with redshift information has traditionally been hampered by the need to image the region around the radio source, preferably at near-infrared wavelengths, and then target the most probable host galaxies with time-intensive spectroscopic observations. Such observational campaigns, which took many years to complete, are now becoming less critical due to the plethora of large and deep imaging surveys covering optical and near-infrared wavebands \citep[e.g.][]{SDSSDR7,WISE,KIDS,Banerji2015}.

Moreover, previous radio surveys were predominantly conducted in the northern hemisphere, as such the wealth of southern hemisphere facilities, such as the Atacama Large Millimetre Array (ALMA), cannot easily be used to obtain more detailed information about the physics at play.

Given the importance of understanding the fuelling of and feedback from powerful radio galaxies, one of the key observations that can be made is that of the neutral and molecular gas in and around the AGN host galaxy. There are several ways of probing this gas, but the most powerful facilities are based in the southern hemisphere. Namely the Atacama Large Millimetre Array (ALMA) operating in the millimetre waveband, the Meer Karoo Array Telescope \citep[MeerKAT; ][]{Jonas2009,Camilo2018} and the Australian Square Kilometre Array Pathfinder \citep[ASKAP;][]{Johnston2008} operating at or below 1.4~GHz, allowing the observation of neutral hydrogen in both absorption \citep[e.g.][]{Allison2015, MALS,MIGHTEE} and emission \citep[e.g.][]{ASKAP-HI, LADUMA,MIGHTEE}.

The aim of this study is to cross-match radio and near-infrared (NIR) sources to produce a catalogue of powerful radio sources for absorption line studies with MeerKAT and other radio facilities.


In this study we cross-match radio detected sources from  the Sydney University Molonglo Sky Survey \citep[SUMSS; ][]{SUMSS1,SUMSS2}  with NIR identifications from the VISTA Hemisphere Survey \citep[VHS; ][]{VHS}, and then use the Dark Energy Survey \citep[DES;][]{DES} optical data to obtain photometric redshifts.  The paper is organised as follows: In Section~2 we provide a description of the three
surveys used to define the radio galaxy sample. In Section~3 we discuss how we use the likelihood ratio technique to define a robust sample of radio galaxies. In Section~4 we describe the photometric redshifts and in Section~5 we discuss our results on the stellar masses and star-formation rates for the radio galaxies in our sample. We conclude in Section~6 and discuss future planned extensions to the work presented here.

Throughout this work we have assumed {\em Planck} cosmology with $H_0 = 67.7$\,km\,s$^{-1}$\,Mpc$^{-1}$,  $\Omega_{\rm M} = 0.31$ and $\Omega_{\Lambda} = 0.69$. We use AB magnitudes unless stated otherwise.

\section{Sample Selection}

In this section we describe the multi-wavelength data we use to define our new sample of powerful radio sources.


\subsection{SUMSS}
In order to create a new sample of powerful radio galaxies in the southern hemisphere, we first require a parent radio survey which covers enough sky area that it provides a statistically significant number of bright radio sources.
The parent radio survey that we use in this study is therefore SUMSS. SUMSS surveyed the sky below $\delta<-30^{\circ}$ with $|b|>10^{\circ}$, at a frequency of
843~MHz using the Molonglo Observatory Synthesis Telescope \citep[MOST;][]{Mills81}. The survey has a resolution of $45''$~cosec$|\delta|\times 45"$ and a 5$\sigma$
sensitivity of 6~mJy/beam at $\delta < -50^{\circ}$ and 10~mJy/beam at $\delta>-50^{\circ}$. SUMSS is the deepest, wide-area radio survey of the Southern sky to
date \citep{SUMSS2}, although this will change with the full operations of ASKAP with the Evolutionary Map of the Universe \citep[EMU; ][]{EMU}. SUMSS data products include 4.3$^{\circ}\times 4.3^{\circ}$ mosaic images and a source catalogue made by fitting elliptical Gaussians to 
5$\sigma$ peaks in the images.\\

We select sources from the SUMSS catalogue detected at $>10\sigma$, i.e. a limiting peak
brightness of 12\,mJy/beam at $\delta \leq-50^{\circ}$ and 20~mJy/beam at $\delta>-50^{\circ}$. The 10$\sigma$ limit ensures a robust sample of radio sources with
good positional accuracy and below this limit the sample becomes incomplete. Using these selection criteria we find $5,380$ radio sources.

\subsection{VISTA Hemisphere Survey (VHS)}

The Visible and Infrared Survey Telescope for Astronomy (VISTA) Hemisphere Survey (VHS) is a near-infrared photometric survey of the southern sky. It covers 18,000\,sq. deg with magnitude limits of $H$ = 19.8; $J$ = 20.6; $K_{\rm s}$ = 18.5 (Vega). The VHS sky coverage is divided into 3 regions:

1. VHS-NGC (North Galactic Cap): $b>30, \delta<0^{\circ}$, covering 2500\,deg$^{2}$ with baseline exposures of 60\,secs per band in $J,H$ and $K_{\rm s}$.

2. VHS-SGC (South Galactic Cap): $b<-30,\delta<0^{\circ}$, covering 8000\,deg$^{2}$ with baseline exposure of 60\,secs per band in $J,H$ and $K_{\rm s}$.

3. VHS-GPS (Galactic Plane Survey): $5<|b|<30^{\circ}$, covering 8500\,deg$^{2}$ with baseline exposures of 60\,secs per band in $J$ and $K_{\rm s}$.

The near-infrared photometric data, which has a resolution of $\sim 1.2$\,arcsec,  was accessed via the VISTA Science Archive\footnote{horus.roe.ac.uk} and we used the VHS Data Release 3.

We used the {\sc mergedclass} source classifier in the VHS catalogue, to separate objects as probable star, probable galaxy, noise or saturated source. 
We applied the flag to the near-infrared source catalogue to select galaxies, retaining all the sources with the {\sc mergedclass}=1. This may remove some radio galaxies from our sample, which may be at lower signal-to-noise ratio and therefore classed as probable galaxies but this is not an issue for our core aims of defining a sample for follow-up studies. Sources that were
saturated, probable star, stars and noise were therefore removed. 

\subsection{Dark Energy Survey (DES)}
The Dark Energy Survey \citep[DES;][]{DES} is an optical survey that will eventually cover 5000\,deg$^{2}$ in the SGP region in the $griz$ bands, using a dedicated camera (DECam) on the 4\,m Victor M. Blanco Telescope at the Cerro Tololo Inter-American Observatory (CTIO) in Chile \citep{Flaugher15}. The DES is aimed at understanding the nature and evolution of dark energy using a multi-probe approach, utilising measurements of large-scale
structure, weak gravitational lensing, galaxy cluster counts and a dedicated multi-epoch supernova survey over 30\,deg$^2$. However, the depth and areal coverage of the DES means it is ideal for identifying rare sources such as radio source host galaxies.

\begin{figure}
\includegraphics[width=\columnwidth]{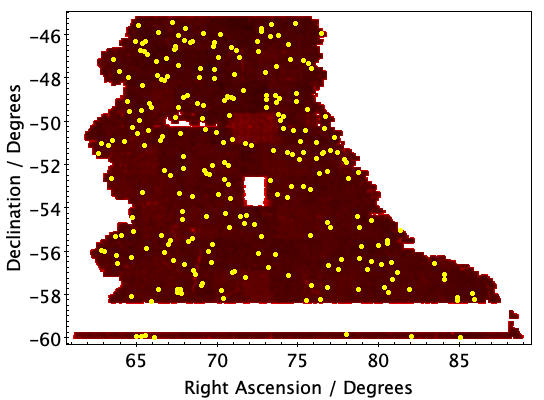}
\caption{Sky coverage of the combined Dark Energy Survey and VISTA Hemisphere Survey DR3 (red background), overlaid with the final cross-matched sample of 249 radio sources (filled yellow circles). The white strip at a declination $\sim$ -59\,degrees and the white rectangle in the middle of the field denote areas that have no near-infrared data from the VHS in DR3.}
\label{fig:coverage}
\end{figure}

We use the DES Science Verification 1 Gold Release\footnote{https://des.ncsa.illinois.edu/releases/sva1} and in the region of overlap with the VHS, we find 9,325,566 sources. The DES also provides photometric redshifts for all of the sources with significant detections.
We remove galaxies that were affected by bad pixels by
rejecting objects with one of the following flags: $Modest\_class \neq$1, $flags\_G >$ 0, $flags\_R >$ 0, $flags\_I >$ 0, $flags\_Z >$ 0, 
$Badflags\_G >$ 0 \citep{Jarvis16}. In this work we take the photometric redshift measurement from the largest portion of the Science Verification (SV) survey. The SV area is approximately 148\,deg$^{2}$ contained within the eastern part of the region observed by the South Pole Telescope (SPTE). The DES optical imaging coverage is shown in Fig.~\ref{fig:coverage} and has a seeing-limited resolution of $\sim 1$\,arcsec.

\section{Likelihood Ratio}

Given the large discrepancy between the resolution of SUMSS and the optical and near-infrared data, we employ the likelihood ratio technique to associate the radio sources, which are predominantly unresolved at the resolution of SUMSS, with the near-infrared catalogue.
The likelihood ratio (LR) is the ratio between the probability that the source is a true identification and the corresponding probability that the source is a
false match \citep[e.g.][]{Sutherland92,Smith11,Fleuren12,Kim12}, and can be expressed as 
\begin{equation}
{\rm LR}=\frac{{q}(m){f(r)}}{{n}(m)} ,
\end{equation}
where $q(m)$ is the true distribution of infrared counterparts (from VHS in this case) as a function of magnitude, $f(r)$ is the radial probability distribution function of
the offset between the SUMSS position and the position of a galaxy in the VHS catalogue, and $n(m)$ is the surface density of galaxies in the VHS survey as a function of magnitude. We use the $K_{\rm s}$-band as the band to perform the likelihood ratio analysis with SUMSS due to the fact that we expect the host galaxies to be massive and near-infrared bright for the more powerful radio sources \citep[e.g.][]{Jarvis2001b,Willott03,Seymour2007}.

The radial probability distribution for infrared counterparts to the SUMSS radio sources is then given by:

\begin{equation}
f(r)=\frac{1}{2\pi \sigma^2_{\rm pos}} {\rm exp} \left(\frac{r^2}{2\sigma^2_{\rm pos}}\right) ,
\end{equation}
where $r$ is the positional offset between the radio and $K_{\rm s}$-band sources, and $\sigma_{\rm pos}$ is the combined positional error of the radio and near-infrared positions. In this study, $\sigma_{\rm pos}$ is dominated by the positional error of the SUMSS radio sources, and we therefore neglect the contribution to this term from the near-infrared sources.
Following \cite{Ivison2007}, we assume that the positional error for the radio sources can be described by

\begin{equation}
\sigma_{\rm pos}=0.655\frac{\rm \theta_{\rm FWHM}}{\rm SNR},
\end{equation}
where $\theta_{\rm FWHM}$ is the full-width half maximum of the synthesised beam, and the SNR is the signal-to-noise ratio for the detected source. We note that the resolution of SUMSS is declination dependent, but our data only span $\sim 15$\,degrees in declination, we therefore take the median resolution of our sample to fix $\theta_{\rm FWHM}$.

\begin{figure}
\includegraphics[width=\columnwidth]{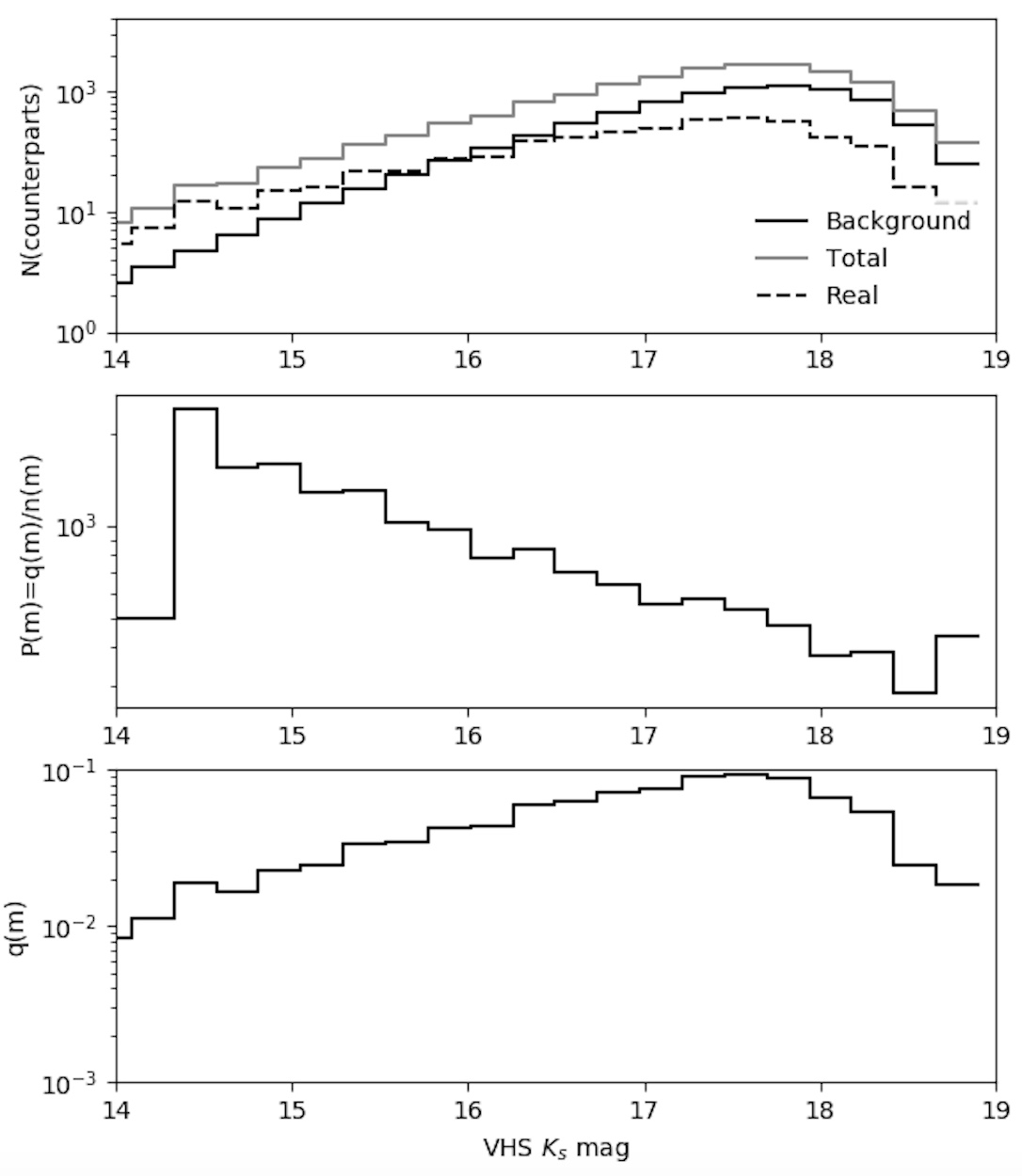}
\caption{{\it (top)} The magnitude distribution of the background galaxy number density, $n(m)$ derived directly from the VHS catalogue, the $total(m)$ denoting all possible matched within radius $r_{\rm max}$, as defined in the text, and the $real(m)$ distribution of actual counterparts. {\it (middle)} shows the magnitude dependent $q(m)/n(m)$ distribution calculated from the data as described. {\it (bottom)} the true distribution, $q(m)$ of the near-infrared counterparts.}
\label{qmdistribution}
\end{figure}

The source density term $n(m)$, is estimated from the source counts of the input VHS catalogue.
We can estimate $q(m)$ following \cite{Ciliegi03} and \cite{Fleuren12}. First we calculate the magnitude distribution, $total(m)$ for all near-infrared objects within a fixed search radius, $r_{\rm max}$ of the centroid of the radio source. The real distribution of counterparts, $real(m)$ can then be determined by subtracting the background number density $n(m)$ normalised to the area defined by $r_{\rm max}$,

\begin{equation}
{real}(m)={{total}(m)}-\left[n(m)\,N_{\rm radio}\,\pi\,r^2_{\rm max}\right] ,
\end{equation}
where $N_{\rm radio}$ is the total number of radio sources in the input catalogue (see 
Figure~\ref{qmdistribution}). The magnitude distribution of all the possible counterparts within $r_{\rm max}$ of the radio positions is then given by,
\begin{equation}
q(m) = \frac{{\rm real}(m)}{\sum_m {\rm real}(m)}\,Q_0  ,
\end{equation}
where $Q_0$ is an estimate of the fraction of SUMSS radio sources with infrared counterparts above the magnitude limit of VHS.



\begin{figure}
\includegraphics[width=\columnwidth]{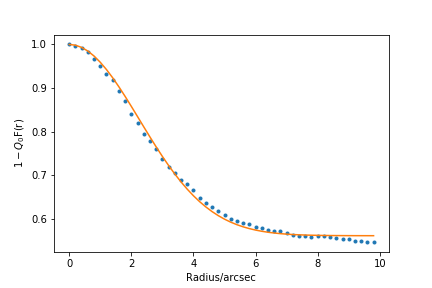}
\caption{Method to estimate $1 - Q_0$ . The solid line is the best fit to the model $1 - Q_0F(r)$, with $Q_0 = 0.395$. The points represent the data obtained by dividing the
number of sources in the SUMSS catalogue without a near-infrared match by the number random positions that didn't have a near-infrared counterpart within radius $r$, i.e. $U_{\rm obs}(r)/ U_{\rm random}(r)$.}
\label{q0}
\end{figure}

To determine $Q_{0}$ we need to determine the number of sources in the radio catalogue with no near-infrared counterparts within a search radius $r_{\rm max}$. We again adopt the method of \citep{Fleuren12}. The number of sources in the radio catalogue without a near-infrared match $U_{\rm obs}$, and the number without a near-infrared identification for a random position within the survey area $U_{\rm random}$, can be related to $Q_0$ via

\begin{equation}
\frac{U_{\rm obs}(r)}{U_{\rm random}(r)} = 1 - Q_0F(r) ,
\end{equation} 
where
\begin{equation}
F(r) = 1 - \exp\left(-\frac{r^2}{2\sigma_{\rm pos}^2}\right).
\end{equation}
Using the observed ratio of $U_{\rm obs}(r)$ to $U_{\rm random}(r)$, we find $Q_0 = 0.395$. Figure~\ref{q0} shows the best fit model compared to the data.

The reliability of the counterparts can then be calculated using,

\begin{equation}
{\rm Rel}_i = \frac{{\rm LR}_i}{\sum_j {\rm {LR}}_j + (1 - Q_0)},
\end{equation} 
where $i$ is the index for the radio sources and $j$ is the index of all the possible near-infrared counterparts.

Similarly, the number of sources with a false identification $N_{\rm cont}$ can be expressed as 

\begin{equation}
N_{\rm cont} = \sum_{\rm {Rel}>0.8}(1-\rm {Rel}).
\end{equation}

The likelihood ratio technique was used to identify the near-infrared counterparts to all the SUMSS radio sources in the overlap area. We ensured that our search radius includes
all possible real counterparts to the radio sources by setting $r_{max}$ to 5 times the largest expected positional error, $\sigma_{\rm pos}=30$~arcsec. Of the 5,380 radio sources above 10$\sigma$ in our radio sample, we find that 1,195 have a near-infrared counterpart with a reliability of Rel$>0.8$, following the threshold used in previous studies using this technique \citep[e.g.][]{Smith11,Fleuren12}. Thus, the majority of radio sources do not have a near-infrared counterpart at the depth of the VHS data. This is in line with expectationsl given the low-resolution radio data and the relatively shallow depth of the near-infrared data \citep[see e.g. Table~2 in ][]{Kim12}.
Furthermore, we expect 81 sources from our sample to have a false identification, corresponding to a contamination of $6.8$ per cent.

It is also worth noting that the resolution of SUMSS means that we cannot disentangle what could be multiple sources within the restoring beam. This is likely to increase the number of radio sources without near-infrared counterparts using the likelihood ratio technique, as the centroid of the radio source may not represent the position of the core associated with the host galaxy. Instead the peak of the emission will have a flux-weighted position that is dependent on the source structure, the relative brightness of any lobe emission and whether there is indeed a strong core at all. Thus, we emphasise that our final sample is not complete but should be representative of the powerful radio galaxy population out to $z\sim 1$, beyond which our optical and near-infrared data are too shallow to consistently detect counterparts to the radio sources.

\section{Photometric redshifts} \label{sec:photoz}

\begin{figure}
\includegraphics[width=\columnwidth]{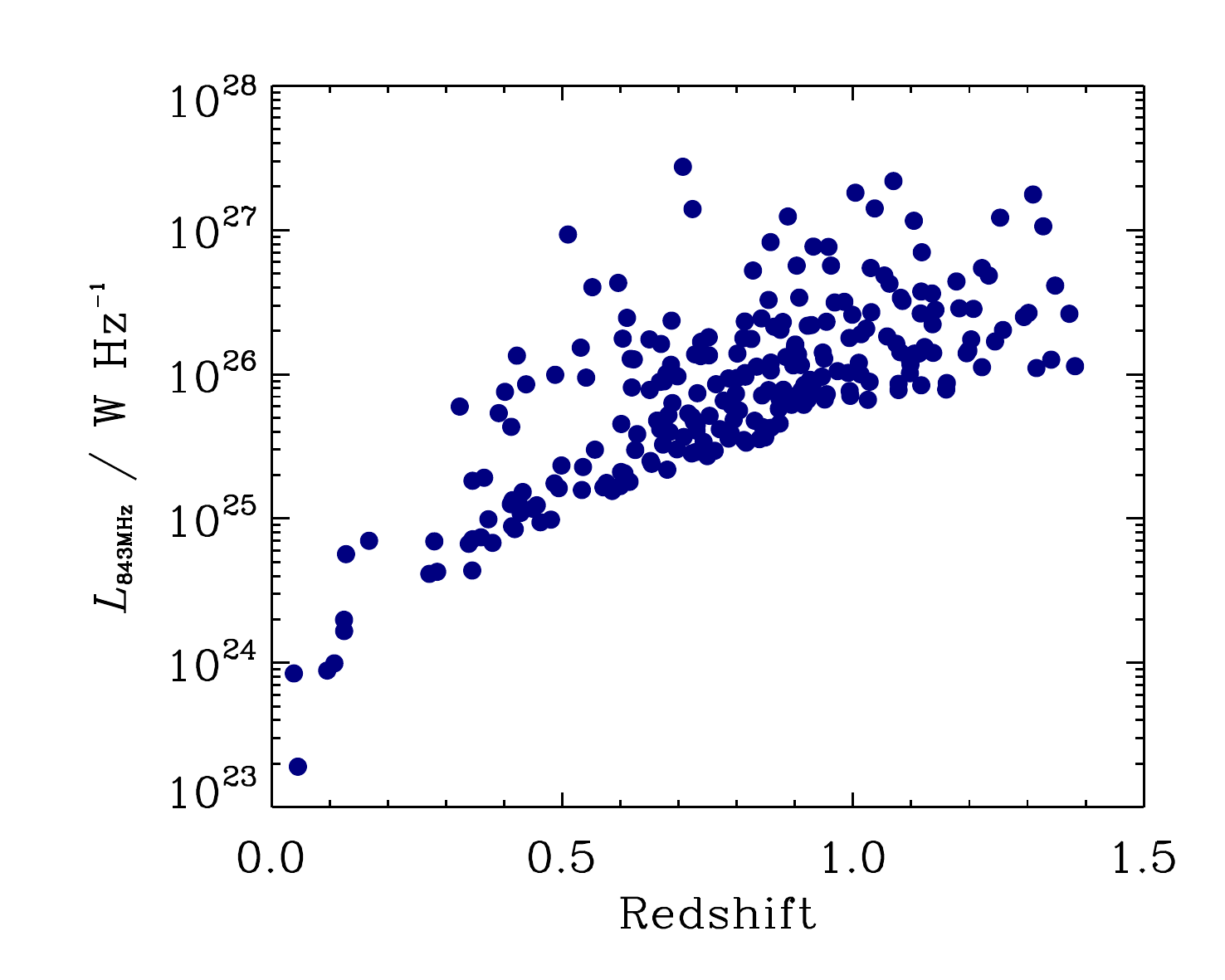}
\caption{Rest-frame 843\,MHz radio luminosity (calculated assuming a spectral index of $\alpha =0.7$) versus redshift for radio galaxies samples from our SUMSS+DES+VHS sample.}
\label{luminosityr}
\end{figure}

Photometric redshifts are an estimate of the redshift of an astronomical object using multi-band photometry instead
of spectroscopy. This technique relies on identifying the passage of continuum features within the spectral energy distribution (SED) across a series of photometric
passbands. Photometric redshifts are essential in the $1<z<2$ range where it is difficult to measure redshifts from optical
spectroscopy due to absence of strong, accessible emission or absorption features. Photometric redshifts are also crucial for detecting galaxies that are fainter than the spectroscopic limits, and
useful for obtaining the redshifts for large numbers of galaxies.  

In this paper we use the DES data products from the first annual reduction of the science verification images (SVA1) GOLD catalogue \citep{Jarvis16,Crocce2016}. The SVA1 GOLD catalogue has
25,227,559 sources with photometric redshifts estimated using Bayesian Photometric Redshift Estimation\citep[BPZ; ][]{BPz}. Spectroscopic redshifts are not available for our radio galaxy sample, however the DES photometric redshift accuracy and how they were estimated can be found in \cite{Sanchez2014}.

In order to obtain the redshifts for our sample, we cross-matched the radio sources with identifications in VHS to the DES catalogue with a search radius of 1\,arcsec. Since the VHS and
DES surveys have high positional accuracy, this 1\,arcsec radius should be sufficient to locate the VHS counterpart in DES. 

Of the 1,195 DES+VHS sources cross-matched to the SUMSS data, we find that 249 have good photometric information that allow the determination of photometric redshifts and these form the sample used from hereon. These are source overlaid on the coverage map in Fig.~\ref{fig:coverage}. We note that the sample is broadly restricted to $z<1$, as the 4000\AA\, break, which is a key spectral feature for fitting photometric redshifts, is redshifted into the near-infrared filters for $z>1$. However, one still requires relatively deep data blueward of the 4000\AA\, break for the photometric redshifts to be accurate. As the main science goal for this sample will be to conduct follow-up studies of the neutral hydrogen, this approximate redshift limit will not adversely effect this goal for current facilities.
Thus, we restrict our sample to the DES-only derived photometric redshifts, but use the near-infrared data when performing spectral energy distribution fitting. Fig.~\ref{luminosityr} shows the photometric redshift versus radio luminosity for our final sample, where we assume a spectral index of $\alpha = 0.7$ to the rest-frame 843\,MHz, where $S_{\nu} \propto \nu^{-\alpha}$.

\begin{figure}
\includegraphics[width=\columnwidth]{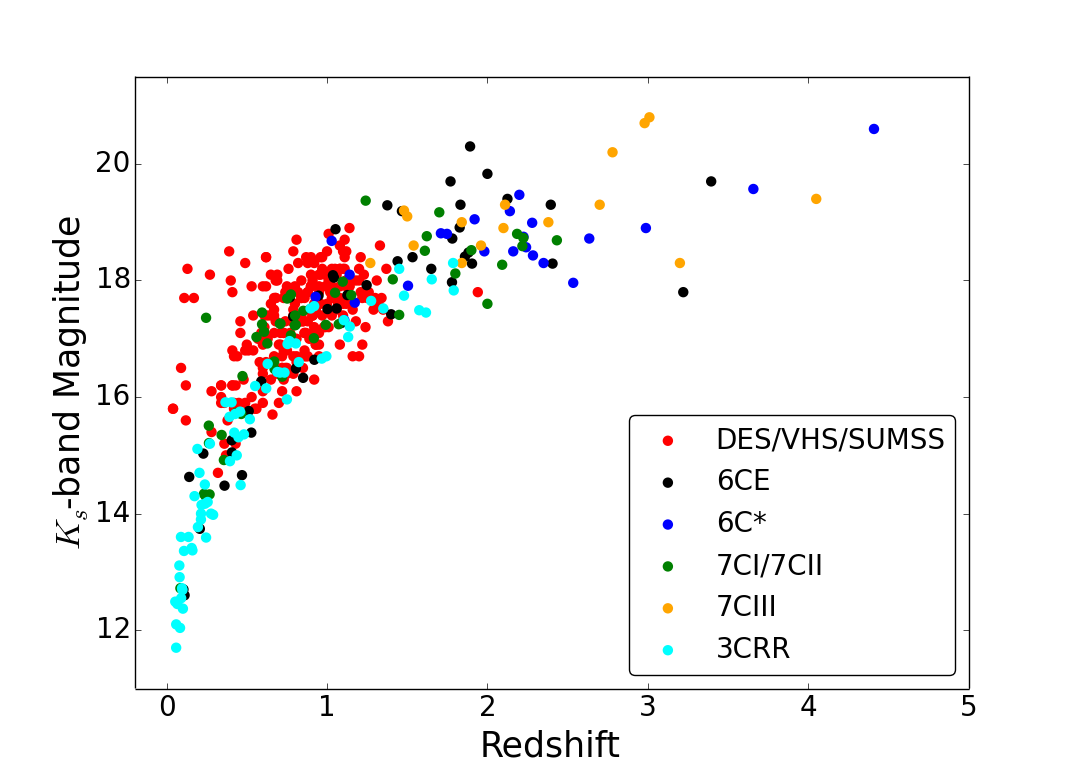}
\caption{$K$-band (Vega) magnitude versus redshift for radio galaxies from our SUMSS/DES/VHS (using the $K_{\rm s}$-band) catalogue (red circles). We also show the points from the 6CE, 6C*,7C-I,II,III and 3CRR complete samples, with colours denoted in the legend.}
\label{k-zrelation}
\end{figure}

\section{The K-z relation}

The study of high redshift radio galaxies can give us a useful insight to the formation of massive galaxies. Many previous studies of the properties of radio galaxy hosts have utilised the so-called $K-z$ relation, which has traditionally been used to demonstrate that powerful radio galaxies tend to reside in the most massive galaxies at all epochs \citep[e.g.][]{Eales1997,Jarvis2001b,Willott03,Rocca04}. In Figure~\ref{k-zrelation}, we show the $K-z$ relation for the radio galaxies in our joint SUMSS+DES+VHS sample, using the photometric redshifts. It is clear that the radio galaxies in our sample broadly follow the $K-z$ relation based on a range of complete samples of radio sources from 3CRR, 6CE, 6C* and 7C, although we note that there is evidence that some of the lower-redshift radio sources in our sample tend to have fainter hosts. This could be due to a relation between radio luminosity and host galaxy mass \citep[e.g.][]{McLureetal2004}, given that our radio sources are derived from a deeper parent radio survey \citep[10\,mJy at 843\,MHz compared to 50\,mJy at 151\,MHz in the case of 7CRS; ][]{Willott03}.  However, at these low redshifts and flux-density limit, we are close to the radio luminosity where star-forming galaxies may be being included \citep[$L_{1.4 \rm GHz} >10^{23}$\,W\,Hz$^{-1}$; ][]{Mauch07} in the sample selection, and we would expect such galaxies to show a marked difference in their host galaxy properties compared with the massive ellipticals that tend to host more powerful radio sources. Another possibility is that these are misidentifications in our likelihood ratio analysis, as we expect 6.8 per cent from our total near-infrared sample to be contaminants. After cross-matching with DES, we may expect this fraction to decrease as we are biased towards brighter objects that are detected at high signal-to-noise at the visible as well as near-infrared wavelengths. However, we would still expect some fraction of our sources to be false associations.  We explore this further in Section~\ref{sec:SFGs}.

\section{Stellar masses of radio galaxies}\label{sec:masses}

Unlike many past studies of radio galaxy host galaxies, given the multiband photometry we are not limited to just using the $K_{\rm s}-$band magnitude in order to compare the radio galaxy hosts to the general galaxy population. We can use the optical photometry from the DES and near-infrared photometry from the VHS to fit stellar population synthesis models in order to obtain estimates of the stellar mass and star-formation rates of the radio galaxy hosts. We therefore use the Fitting and Assessment of Synthetic Templates \cite[FAST; ][]{Kriek2009} code to find the best-fit galaxy templates to the combination of the DES and VHS photometry, using the photometric redshifts described in Section~\ref{sec:photoz}.

We ran the FAST code using two different stellar population synthesis models (SPS) from \citet{BruzualCharlot2003} and \citet{Maraston2005}, with both fixed metallicity and allowing the metallicity to vary, and with a total dust extinction in the range $0.0 < A_V < 0.3$. For this study we assumed an exponentially declining star formation history (SFH) in all different models with a Salpeter Initial Mass Function \citep[IMF; ][]{Salpeter1955}. We also investigated how different choices for the SFH and IMF alter our results, and do not find any significant differences for the stellar mass estimates with a scatter in the stellar masses of $\Delta M_{\star} < 0.2$~dex.

\begin{figure}
\includegraphics[width=\columnwidth]{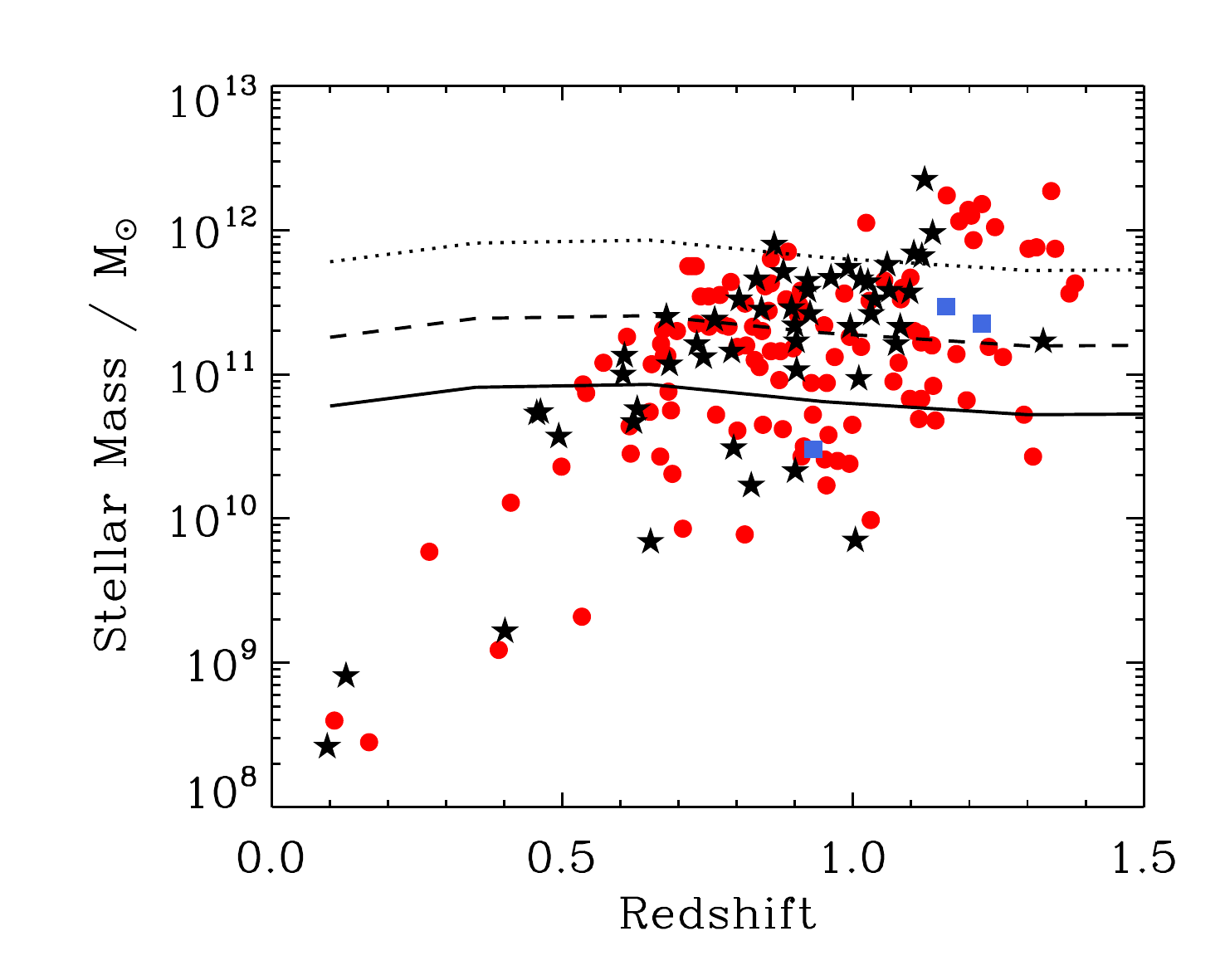}
\caption{Redshift versus stellar mass for radio galaxies in our sample. The red circles denote those radio sources missing $J-$ and $H-$band data, the blue squares are those missing just $J$-band data and the black stars denote those radio  galaxies with a full complement of $grizJHK$ data, demonstrating that the lack of information in some filters does not introduce any obvious systematics in our stellar mass estimates.
  We only show points with SED fits with $\chi^2 <20$. The solid line represents $M_{\star}$ from \citet{Wright2017} at $z<0.1$ and \citet{Ilbert2013} at $z>0.1$. The dashed line corresponds to $5\,M_{\star}$ and the dotted line to $10\,M_{\star}$.}
\label{fig:zmass}
\end{figure}

Figure~\ref{fig:zmass} shows the redshift against stellar mass of the radio galaxy hosts after fitting simple stellar populations to the broad band photometry using FAST. We find a broad range in stellar masses, but with the bulk of the high redshift radio source hosts lying between stellar masses of $10^{11} - 10^{12}$~M$_{\odot}$. This is consistent with the range in stellar masses for the much more radio luminous radio sources that have been studied in detail from surveys such as 3CRR, 6C and 7CRS amongst others. For example, \citet{Seymour2007} undertook a study of the host galaxies of a large sample of powerful radio galaxies using {\em Spitzer} and found host galaxy masses of $10^{11} - 10^{11.5}$~M$_{\odot}$.  Comparing to the non-active galaxy population, this range corresponds to the most massive subset of galaxies in the Universe, at $1-10$\,M$_{\star}$ using the galaxy mass functions from \citep{Wright2017} and \citep{Ilbert2013}.

Although some studies have found a link between radio luminosity and host galaxy mass \citep[e.g.][]{Jarvis2001b,McLureetal2004}, we find no evidence for this in our sample. However, the fact that we only have only used a single flux-limited radio survey means that the range in radio luminosity for a given redshift range is limited. Combining the sample discussed here with both wider area brighter samples, and fainter samples over smaller areas would allow this question to be investigated in much more detail,  as this would increase the range in radio luminosity that could be probed, across the full redshift range.

\begin{figure}
\includegraphics[width=\columnwidth]{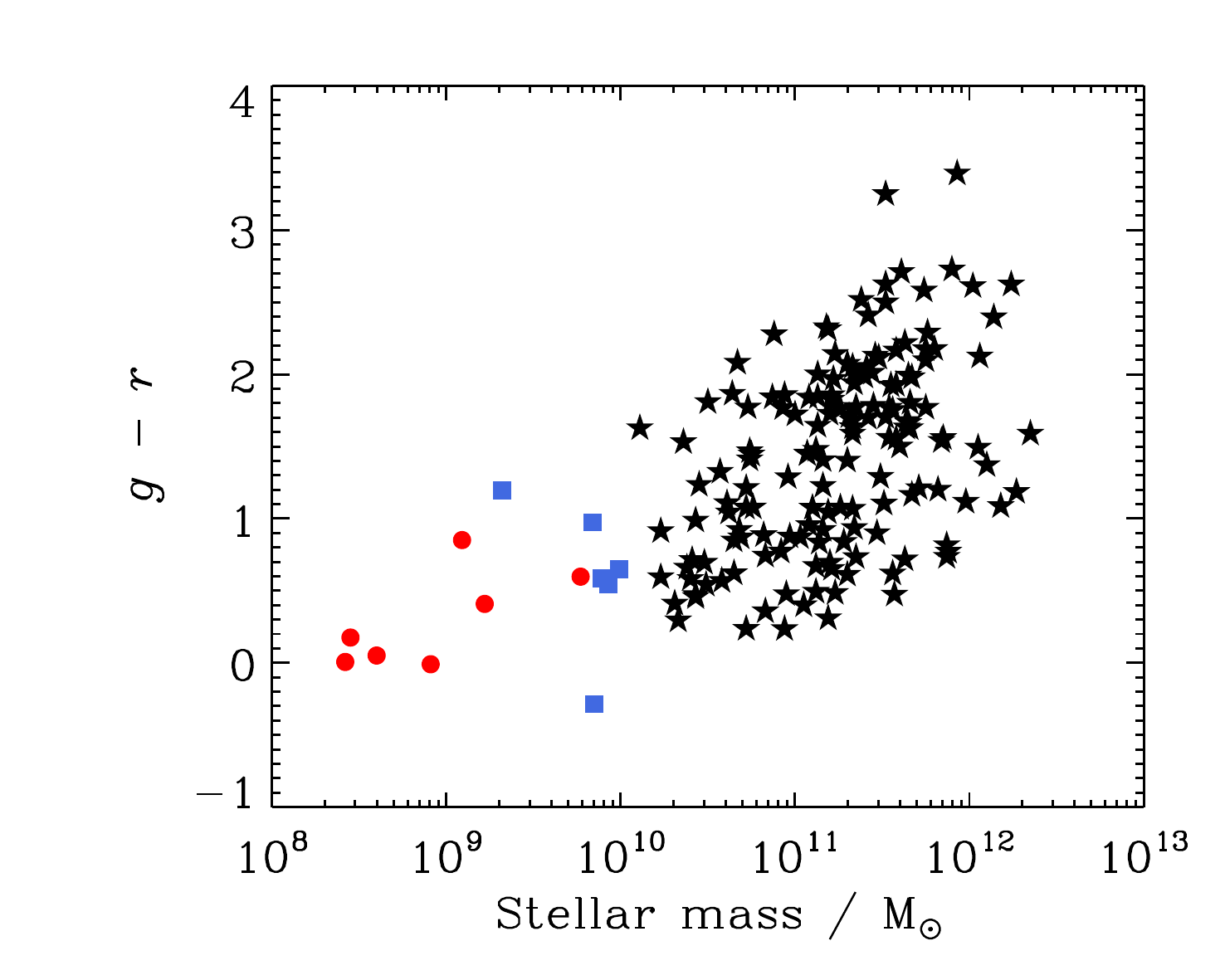}
\caption{Stellar mass versus $g-r$ colour for the radio galaxies in our sample. The red circles denote galaxies at $z<0.5$, whereas the blue squares represent those galaxies at $z>0.5$ with stellar mass of $M_{\star} < 10^{10}$\,M$_{\odot}$. We only show points with SED fits with $\chi^2 <20$.}
\label{fig:colour-mass}
\end{figure}

\section{Star forming galaxies}\label{sec:SFGs}
\begin{figure}
\includegraphics[width=\columnwidth]{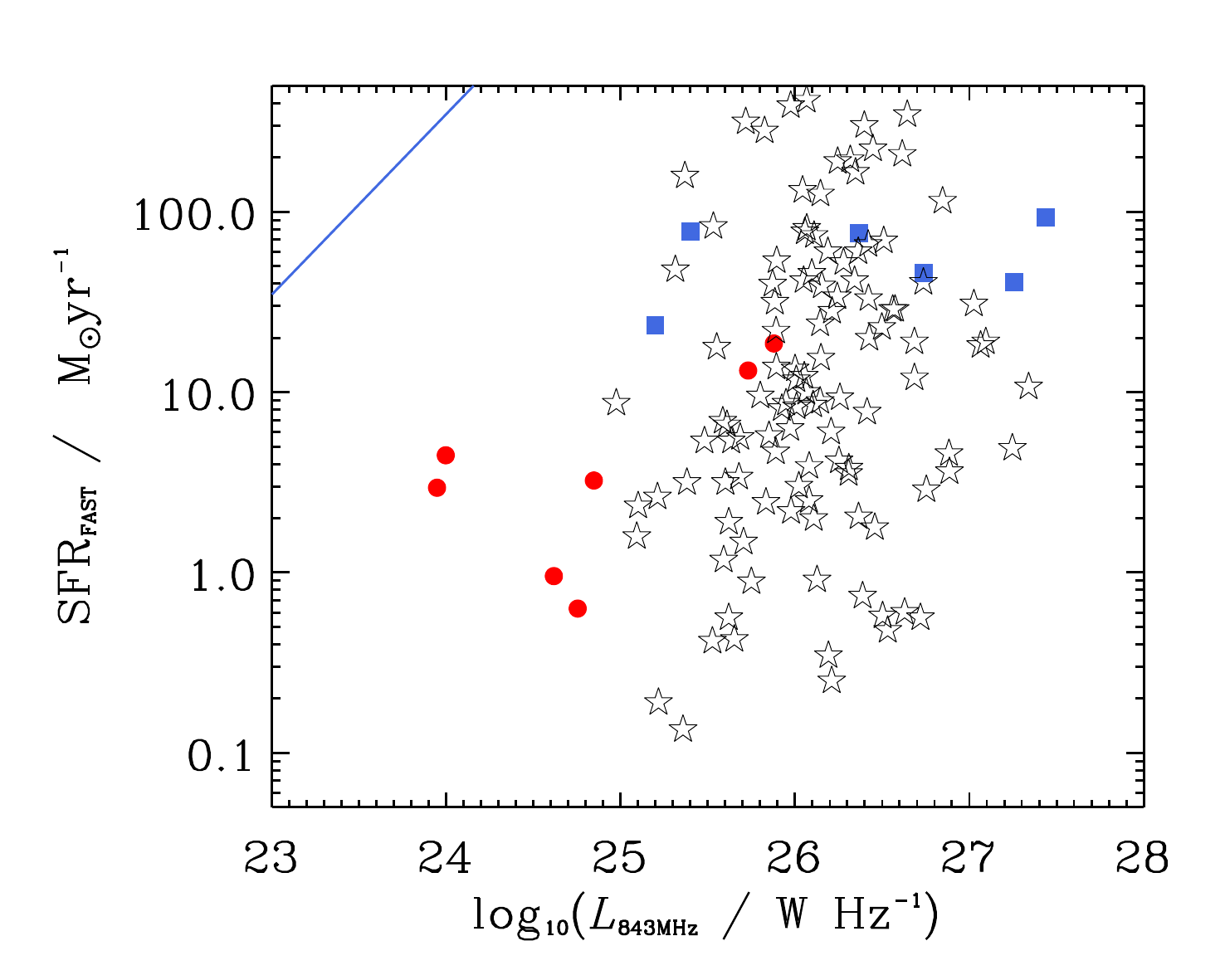}
\caption{Radio luminosity against star-formation rate derived using SED fitting in FAST for the host galaxies of the radio sources in our sample. The red circles denote galaxies at $z<0.5$ with $M < 10^{10}$~M$_{\odot}$, whereas the blue squares represent those galaxies at $z>0.5$ with $M < 10^{10}$~M$_{\odot}$ . The open stars are the galaxies with $M > 10^{10}$~M$_{\odot}$ from the SED fitting. We only show points with SED fits with $\chi^2 <20$. The solid line represents the expected correlation using the relationship between star-formation rate and radio luminosity used in \citet{Delhaize2017} and \citet{Novak2017}.}
\label{fig:sfr-rad}
\end{figure}

There are clearly some radio sources in our sample where the host galaxy is significantly less massive than $\sim 10^{11}\,M_{\odot}$. These are predominantly at low redshift, which is likely due to the depth of the optical and near-infrared data. These galaxies, although relatively bright radio sources, are also much less massive. The majority of these sources tend to be much bluer than the more massive galaxies (Fig.~\ref{fig:colour-mass}). This suggests that these lower-mass, lower-redshift sources are a distinct population from the higher redshift, more powerful sources. Using the radio luminosity function determined by \citet{Mauch07}, the dominant population at $z< 0.3$ $L_{1.4\rm\,GHz} < 10^{23}$\,W\,Hz$^{-1}$ are star-forming galaxies, which corresponds to a radio luminosity of $L_{843\rm\,MHz} = 1.4\times 10^{23}$\,W\,Hz$^{-1}$ at 843~MHz, assuming $\alpha=0.7$. This is less luminous than the radio luminosity of the sources with low stellar mass and blue colours in our sample. However, using the radio luminosity function for star forming galaxies from \cite{Mauch07} we expect to find around 2--3 star-forming galaxies at $z<0.2$ with $L_{843\rm\,MHz} > 1.4 \times 10^{23}$\,W\,Hz$^{-1}$ in our survey area, which is marginally consistent with the number of radio sources with blue colours at these radio luminosities within our sample.
We can use the outputs of the FAST code in order to investigate whether the star-formation rate that we infer from the radio luminosity is consistent with that using the optical and near-infrared photometry. Figure~\ref{fig:sfr-rad} shows the star-formation rate derived from FAST and the 843\,MHz radio luminosity, which clearly shows that the radio power of our low-redshift radio galaxies far exceed that expected from star formation alone. 

\begin{figure*}
\includegraphics[width=0.5\columnwidth]{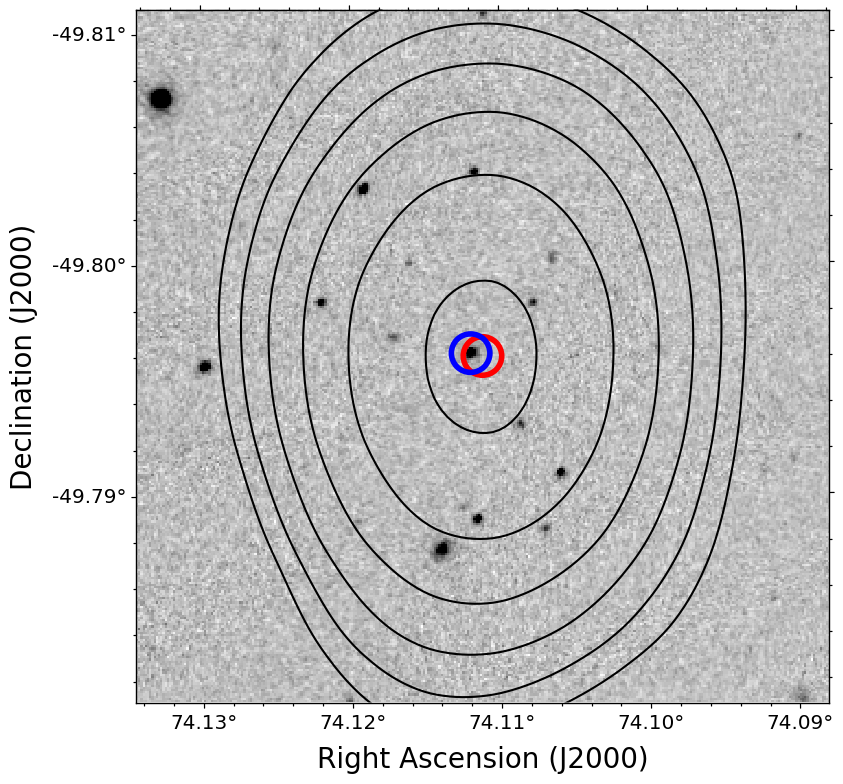}
\includegraphics[width=0.5\columnwidth]{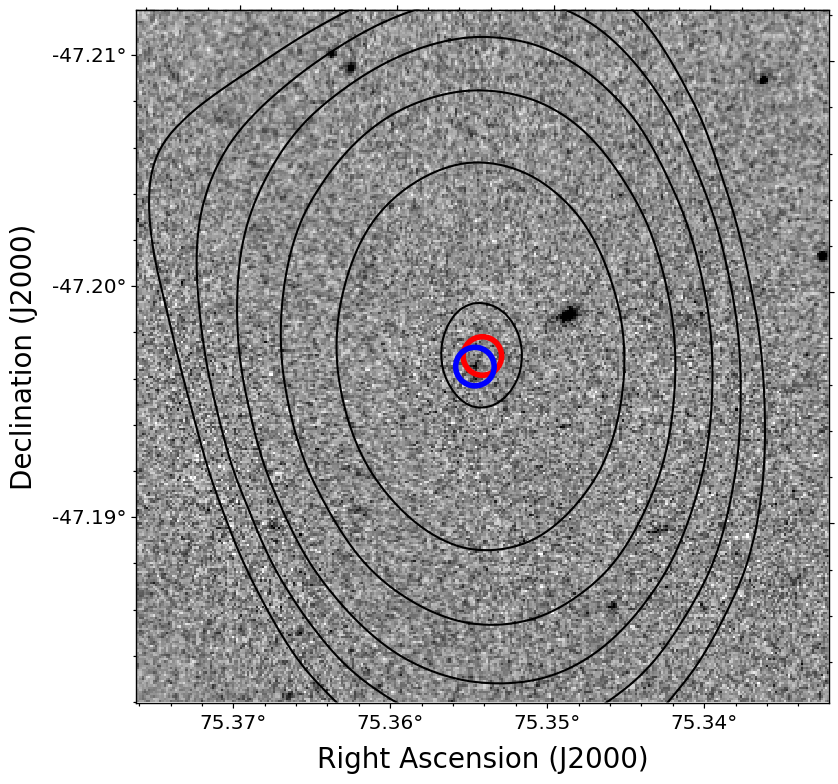}
\includegraphics[width=0.5\columnwidth]{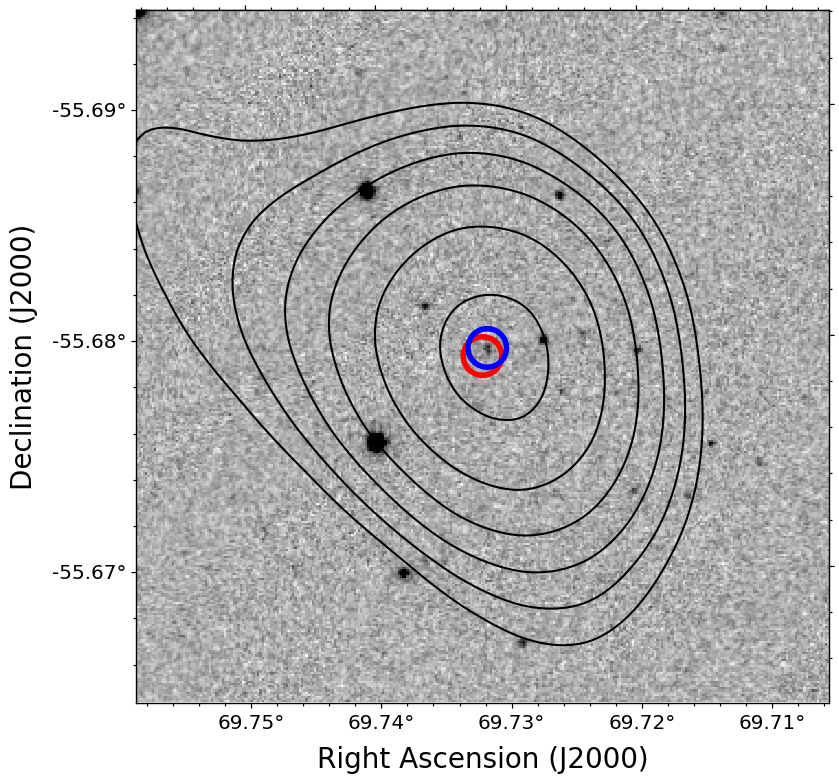}
\includegraphics[width=0.5\columnwidth]{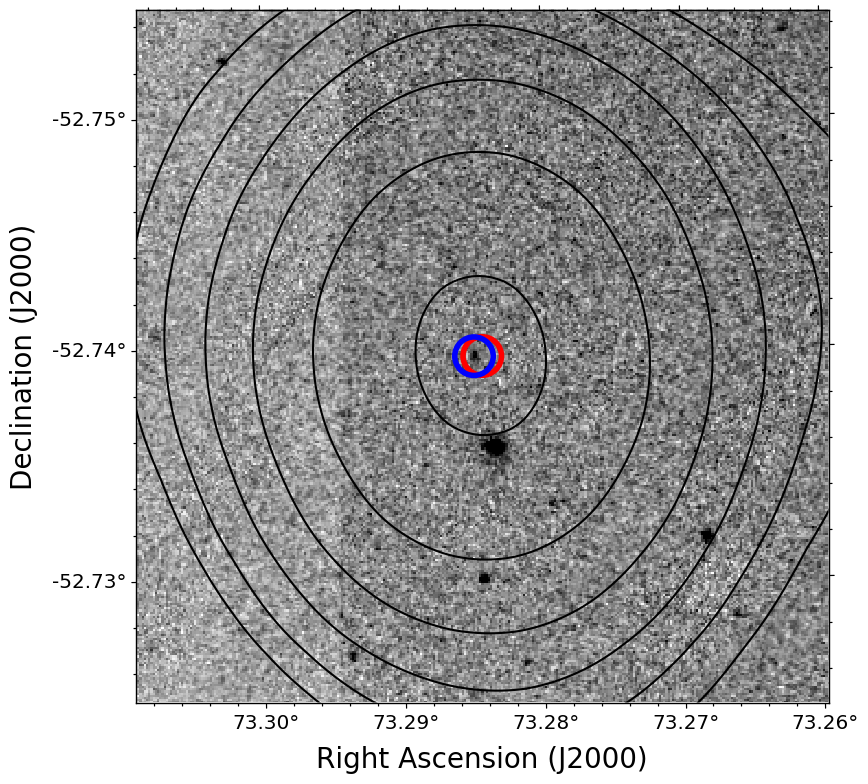}
\includegraphics[width=0.5\columnwidth]{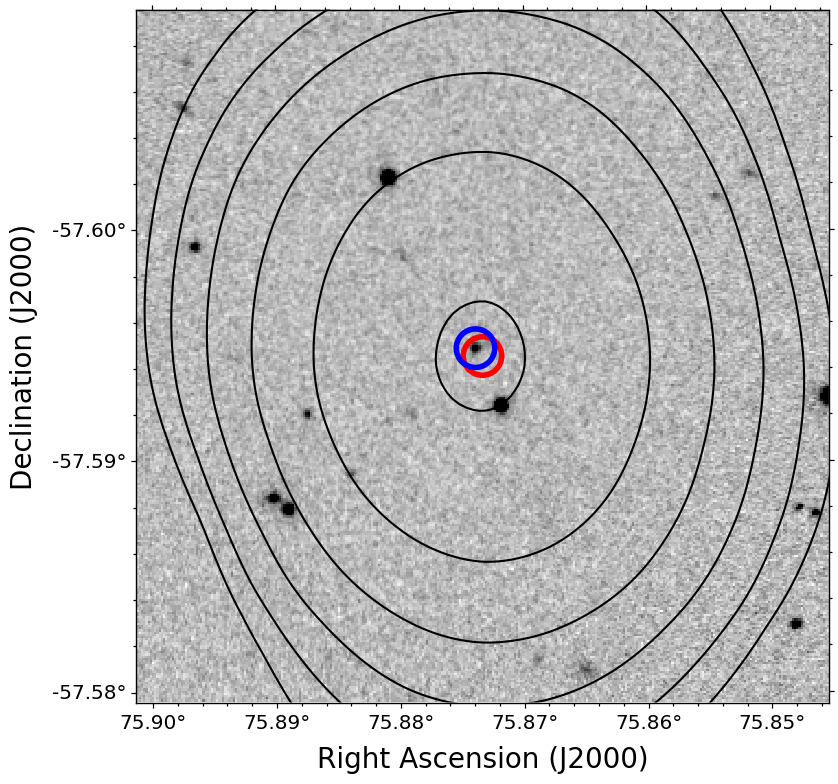}
\includegraphics[width=0.5\columnwidth]{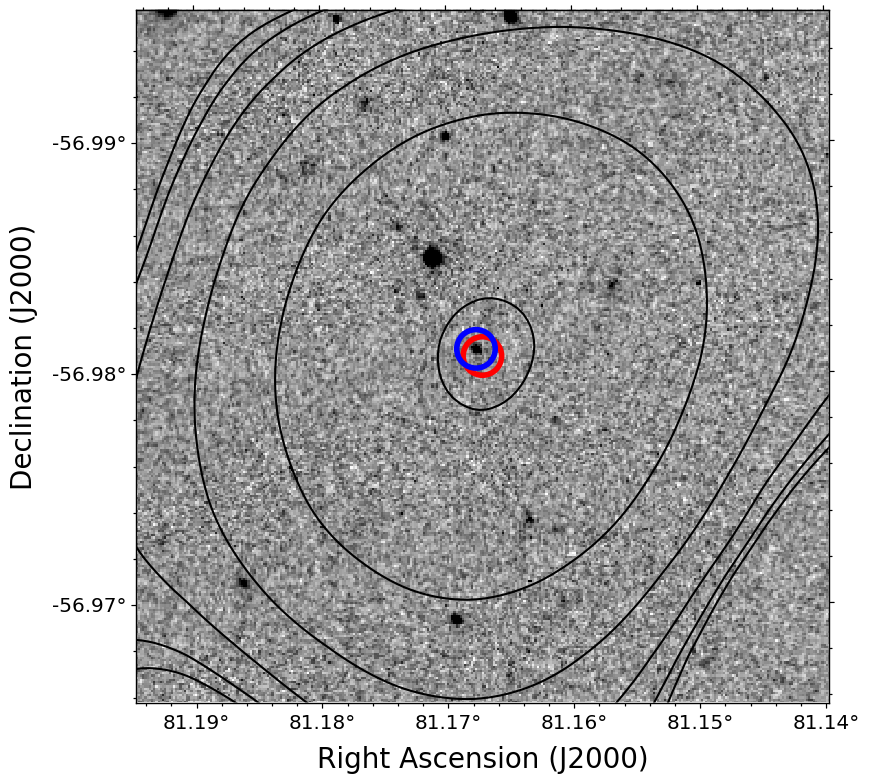}
\includegraphics[width=0.5\columnwidth]{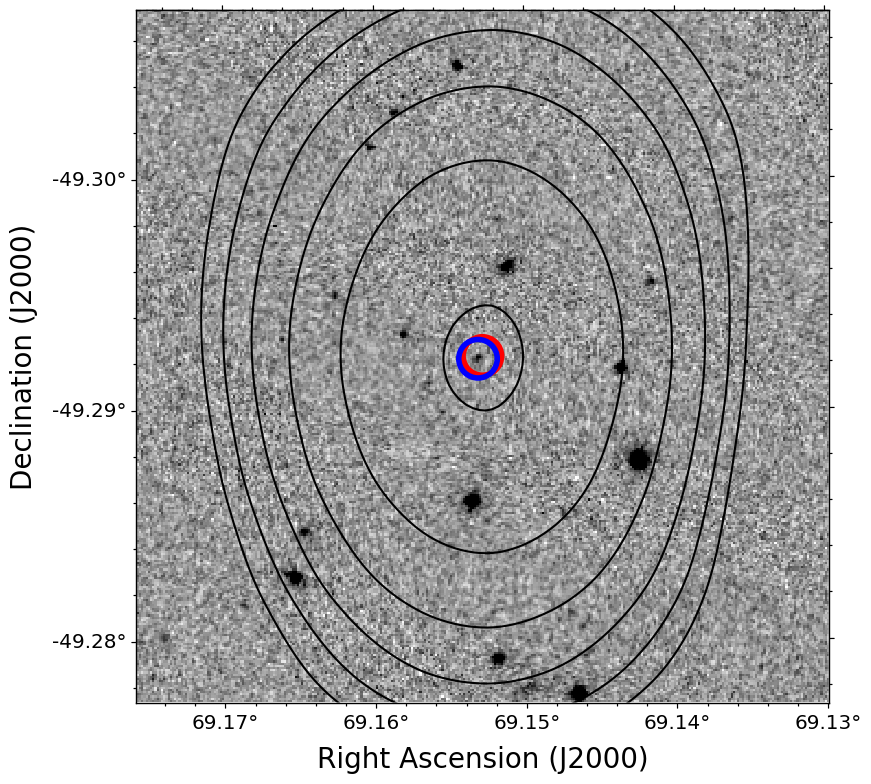}
\includegraphics[width=0.5\columnwidth]{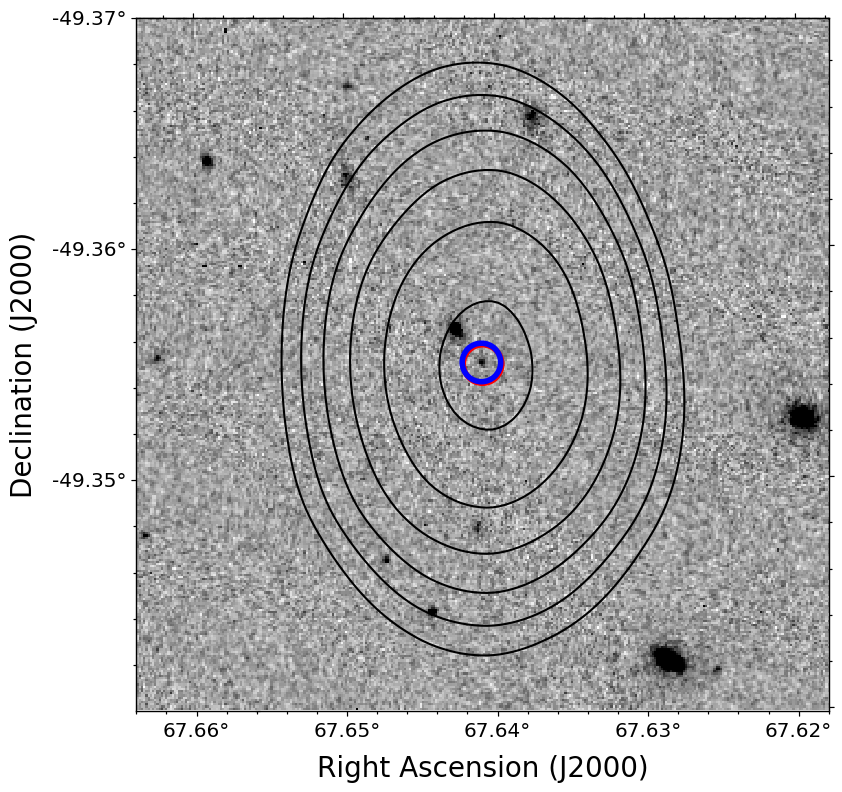}
\includegraphics[width=0.5\columnwidth]{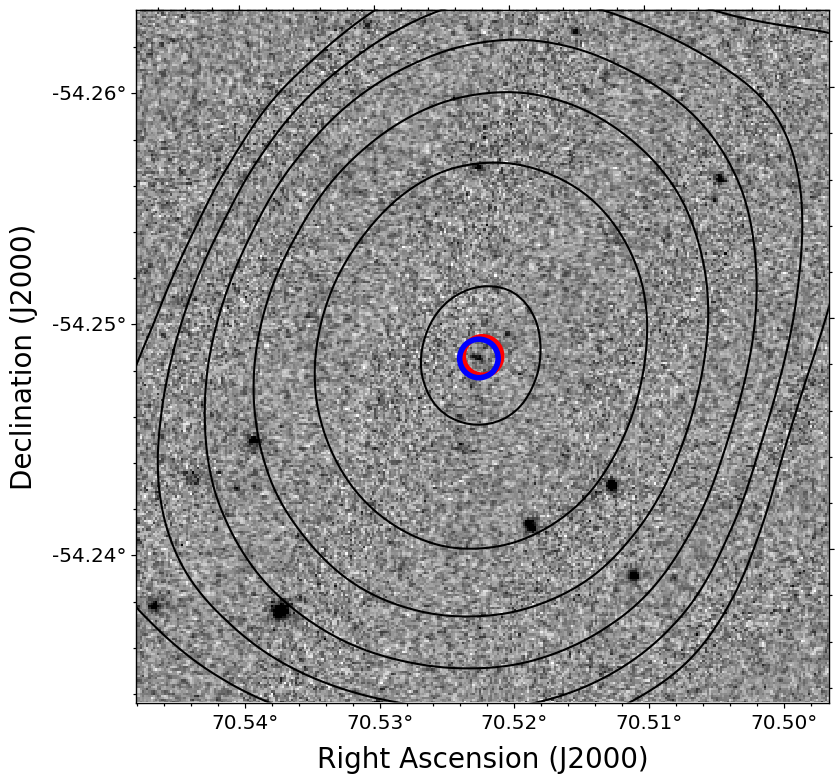}
\includegraphics[width=0.5\columnwidth]{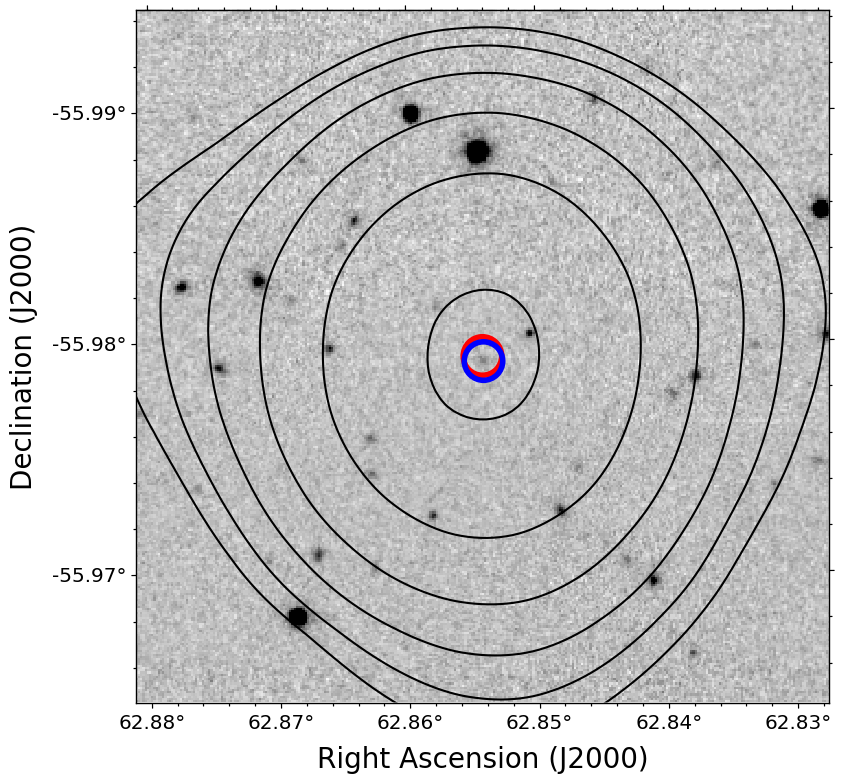}
\includegraphics[width=0.5\columnwidth]{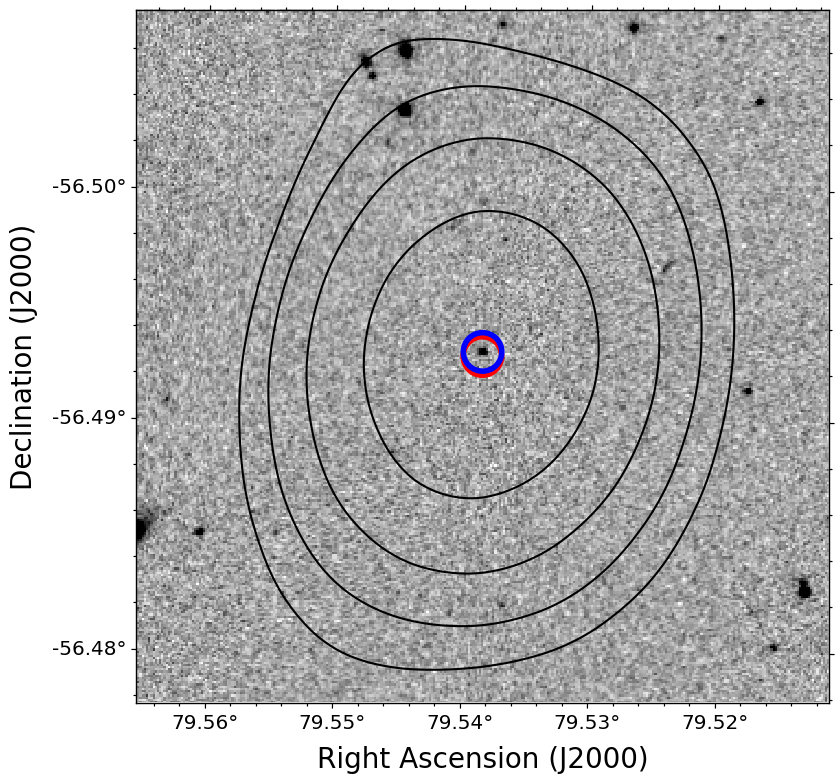}
\includegraphics[width=0.5\columnwidth]{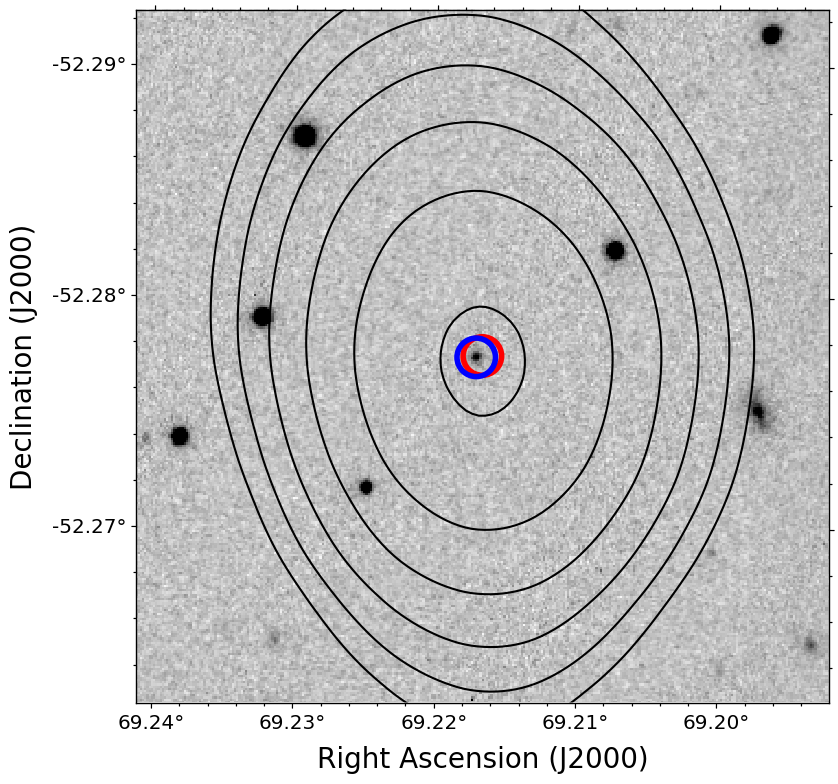}
\includegraphics[width=0.5\columnwidth]{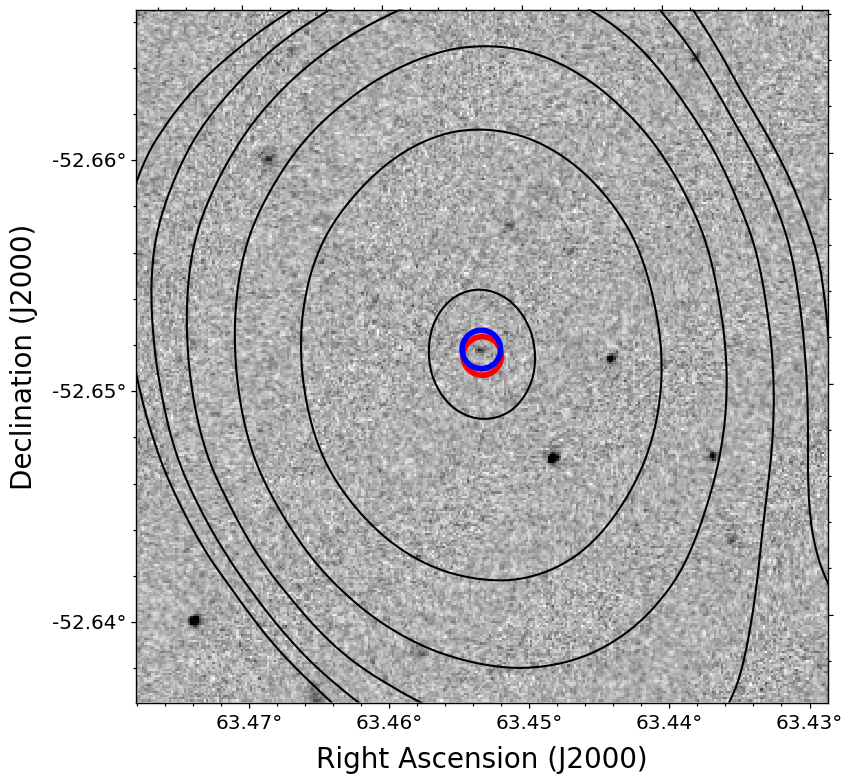}
\includegraphics[width=0.5\columnwidth]{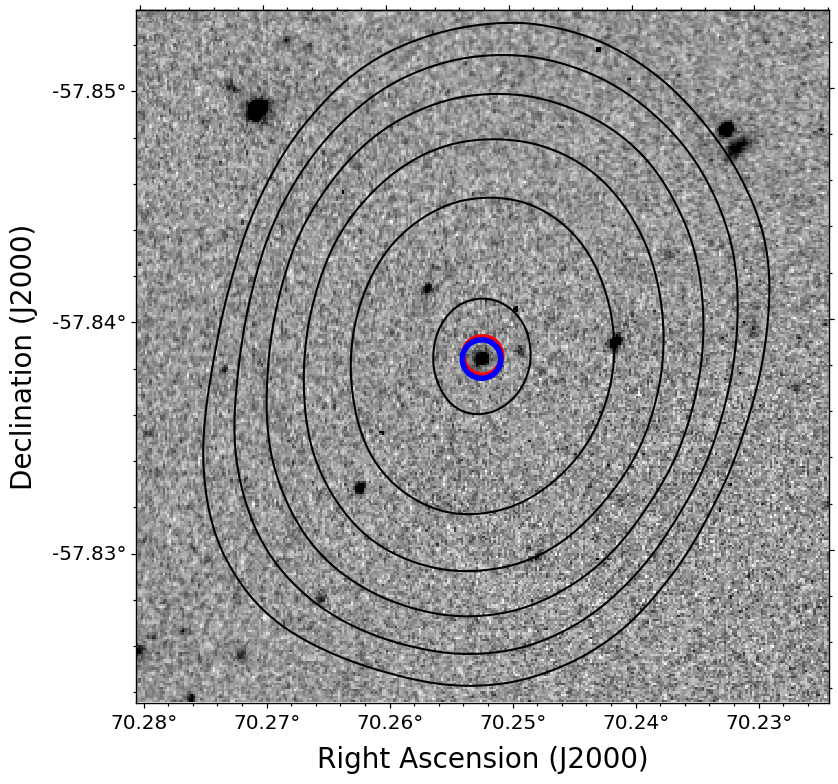}
\includegraphics[width=0.5\columnwidth]{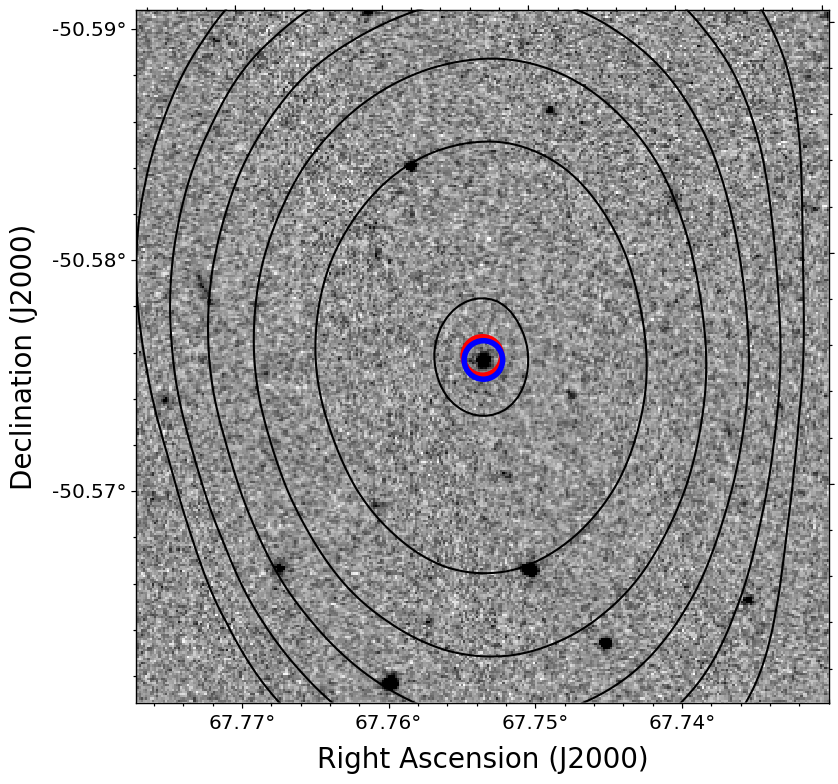}
\caption{Cut-outs of the radio galaxy hosts with estimated stellar mass of $< 10^{10}$~M$_{\odot}$. The sources are ({\em top row})  ID99, 100, 114, 121, ({\em 2nd row}) 130, 136, 159, 183, ({\em 3rd row}) 194, 203, 216, 218, ({\em bottom row}) 220, 243 and 246. The greyscale is the VHS $K_{\rm s}-$band image and the radio contours are from SUMSS, with contours plotted in increments of 4~mJy up to 85 per cent of the peak pixel flux density. The listed radio position is shown by the red circle and the highest likelihood VHS-near-infrared source is dneoted by the blue circle.}
\label{fig:cutouts}
\end{figure*}

We therefore suggest that these sources are more likely to be the small number of low-redshift interlopers that we expect from the estimated false positives in our likelihood ratio analysis. We checked the reliability from the likelihood ratio of these low-mass, low-redshift galaxies and they are not near to the cut-off of 0.8. However, we show the radio contours overlaid on the VHS $K_{\rm s}-$band images for those radio galaxy hosts with an estimated stellar mass of $M_{\star} < 10^{10}$~M$_{\odot}$ in Fig.~\ref{fig:cutouts}. There is no obvious indication of these being misidentified but we cannot rule out fainter high-redshift true counterparts. On the other hand they could be low-mass blue galaxies that host powerful radio sources. Higher resolution radio data and/or deep optical/near-infrared data would be needed to confirm which of these scenarios is correct.

\section{Conclusions}\label{sec:conclusions}
We have described the process of obtaining reliable optical and near-infrared counterparts to a relatively bright sample of radio sources in order to provide a catalogue for follow-up observations with the new generation of radio telescopes. We
found 1195 radio source counterparts with reliability above 0.8, with 81 of these sources expected to be falsely identified. After cross-matching these sources with the Dark Energy Survey photometric redshift catalogue, we form a sample of 249 radio galaxies with redshift information.

Using this sample we investigate the host galaxy properties of these powerful radio galaxies. We find that they follow the well known radio galaxy $K-z$ relation when compared to previous complete samples selected at higher radio flux-density limits. 

We then performed spectral energy distribution fitting to the optical and near-infrared photometry to investigate the host galaxy properties. We found that the radio galaxies fall into two broad populations, low-mass blue galaxies and high-mass red galaxies. The red galaxies, which are dominated by the radio-AGN, typically have host galaxy stellar masses in the range $10^{11} -10^{12}$\,M$_{\odot}$, which corresponds to $1-10$\,M$_{\star}$ from the galaxy mass function, suggesting that in order to produce powerful radio emission, the galaxy needs to be sufficiently massive to host supermassive black hole of $>10^{8}$\,M$_{\odot}$, if the $M_{\rm bulge} - M_{\rm BH}$ relation holds to high redshift. This would be in line with studies of radio-loud quasars, where the black-hole mass is measured via the virial method \citep[e.g.][]{McLure04}.

The lower-mass $M<10^{10}$\,M$_{\odot}$ galaxies in our sample of radio sources are consistent with being blue star-forming galaxies. We compare the star-formation rates derived from SED fitting to those based on the radio luminosity and find that the radio luminosity is significantly higher than one would expect. We therefore suggest that these sources are likely to be made up of the small fraction of high-reliability identifications that are actually false positives. On the other hand they could be lower-mass blue galaxies hosting bright radio sources, or their photometric redshifts could be significantly wrong. Follow up, higher-resolution radio imaging would help alleviate mid-identifications, as the limiting factor in our cross-identifications is the low resolution of the SUMSS radio imaging. Spectroscopy to confirm their redshifts would also be useful to determine their nature.

This sample of southern bright radio sources will be used as a basis for studying the fuelling and feedback from powerful radio galaxies using the new generation of radio telescopes. In particular, we plan to undertake a comprehensive study of H{\sc i} absorption against these powerful sources with the MeerKAT radio telescope.


\begin{landscape}

  \begin{table}
  \caption{The radio galaxy sample overlapping SUMSS+VHS+DES defined in this paper. The photometric redshift uncertainties given are from the 32nd and 68th percentiles of the PDF supplied in the DES SVA Gold data release. The full table is available online.}\label{tab:example_table1}
{\tiny
  \begin{tabular}{lllllllllllllr} 
   \hline
   ID&RA&Dec&$g$&$r$&$i$&$z$&$J$&$H$&$K_{s}$&$S_{843} /$&$z_{\rm phot}$&$\log_{10}(M_{\star} / $&$\log_{10}$(SFR / \\
  & & & & & & & & & &mJy& &M$_{\odot}$)&M$_{\odot}$/yr) \\
\hline
1&04:50:00.67&-46:18:12.9&24.190$\pm$0.159&22.426$\pm$0.040&21.307$\pm$0.026&20.242$\pm$0.017&19.281$\pm$0.159&19.096$\pm$0.183&18.711$\pm$0.182&17.00$\pm$1.00&1.08$^{+0.03}_{-0.04}$&12.04&3.24\\
2&05:10:03.35&-56:51:59.8&21.006$\pm$0.008&20.769$\pm$0.009&20.426$\pm$0.012&20.014$\pm$0.015&19.870$\pm$0.129& &19.328$\pm$0.255&20.70$\pm$1.50&0.93$^{+0.03}_{-0.08}$&10.72&1.60\\
3&05:03:02.42&-55:17:54.9&25.479$\pm$0.488&22.897$\pm$0.063&21.752$\pm$0.041&20.810$\pm$0.035&20.110$\pm$0.199&19.603$\pm$0.209&19.071$\pm$0.231&24.60$\pm$1.00&0.99$^{+0.01}_{-0.03}$&11.74&-2.06\\
4&04:21:14.97&-59:55:37.6&22.399$\pm$0.028&21.922$\pm$0.039&21.364$\pm$0.025&20.786$\pm$0.030&20.008$\pm$0.168& &19.428$\pm$0.189&441$\pm$13&1.07$^{+0.04}_{-0.03}$&10.95&1.03\\
5&04:30:31.14&-46:02:48.6&21.900$\pm$0.025&20.118$\pm$0.006&19.430$\pm$0.006&19.065$\pm$0.008&18.628$\pm$0.091&18.093$\pm$0.099&18.150$\pm$0.120&12.70$\pm$1.00&0.48$^{+0.01}_{-0.01}$&11.17&0.64\\
6&04:39:53.16&-50:31:04.4&25.398$\pm$0.487&23.897$\pm$0.178&23.007$\pm$0.135&21.880$\pm$0.072& & &20.461$\pm$0.328&62.90$\pm$2.10&1.08$^{+0.06}_{-0.05}$&11.60&1.84\\
   7&05:06:05.83&-51:28:35.2&24.312$\pm$0.262&22.320$\pm$0.048&21.192$\pm$0.039&20.342$\pm$0.029&19.828$\pm$0.225&19.402$\pm$0.248&18.984$\pm$0.247&19.00$\pm$0.90&0.92$^{+0.03}_{-0.01}$&11.65&-2.15\\
 \hline
 \end{tabular}
}
\end{table}

\end{landscape}

\section*{Acknowledgements}
TM was supported by the South African Radio Astronomy Observatory (SARAO) and National Research Foundation (Grant No. 75415). MJJ, SF, KM and MP also acknowledge support from SARAO. 

\bibliographystyle{mnras}
\bibliography{paper}

\bsp	

\label{lastpage}
\end{document}